\documentclass{article}

\usepackage{arxiv}

\usepackage[utf8]{inputenc} % allow utf-8 input
\usepackage[T1]{fontenc}    % use 8-bit T1 fonts
\usepackage{hyperref}       % hyperlinks
\usepackage{url}            % simple URL typesetting
\usepackage{booktabs}       % professional-quality tables
\usepackage{amsfonts}       % blackboard math symbols
\usepackage{nicefrac}       % compact symbols for 1/2, etc.
\usepackage{microtype}      % microtypography
\usepackage{lipsum}
\usepackage{graphicx}
\usepackage{comment}
\usepackage{amssymb}
\usepackage{pifont}
\usepackage{subfigure}
\usepackage[dvipsnames]{xcolor}
\usepackage{graphicx}
\usepackage[utf8]{inputenc}
\usepackage[T1]{fontenc}
\usepackage{authblk}

\newcommand{\xmark}{\ding{53}}%

\usepackage{xcolor}
\definecolor{light-gray}{gray}{0.95}
\newcommand{\code}[1]{\colorbox{light-gray}{\texttt{#1}}}

\title{Plato Dialogue System\thanks{\url{https://github.com/uber-research/plato-research-dialogue-system}}: A Flexible conversational AI Research Platform}

\author[]{Alexandros Papangelis}
\author[]{Mahdi Namazifar}
\author[]{Chandra Khatri}
\author[]{Yi-Chia Wang}
\author[]{Piero Molino}
\author[]{Gokhan Tur}
\affil[]{Uber AI}
\affil[]{apapangelis@uber.com}

\begin{document}

\maketitle

\begin{abstract}
As the field of Spoken Dialogue Systems and Conversational AI grows, so does the need for tools and environments that abstract away implementation details in order to expedite the development process, lower the barrier of entry to the field, and offer a common test-bed for new ideas. In this paper, we present Plato, a flexible Conversational AI platform written in Python that supports any kind of conversational agent architecture, from standard architectures to architectures with jointly-trained components, single- or multi-party interactions, and offline or online training of any conversational agent component. Plato has been designed to be easy to understand and debug and is agnostic to the underlying learning frameworks that train each component.
\end{abstract}

% keywords can be removed

\section{Introduction}

\begin{figure}[h]
\includegraphics[width=\textwidth]{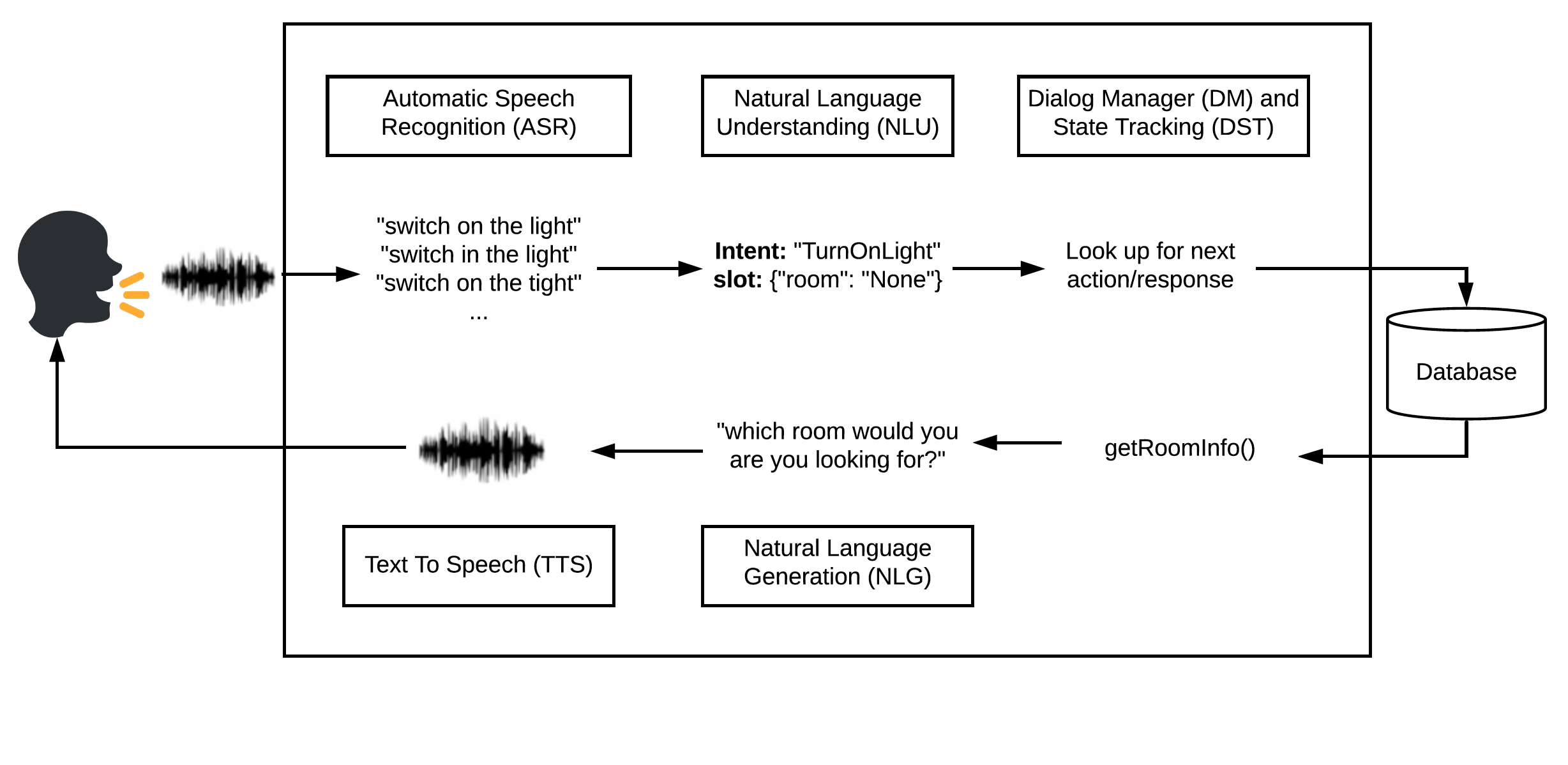}\label{fig:typical_ds}\caption{Depiction of a typical Dialogue System}
\end{figure}

Conversational AI systems traditionally comprise multiple modules such as Automatic Speech Recognition (ASR), Language Understanding (LU), Dialogue State Tracking (DST), Dialogue Management (DM), Language Generation (LG), and Text To Speech (TTS). While a full-fledged spoken dialogue system typically incorporates all these components (depicted in Figure \ref{fig:typical_ds}), most research studies have been performed in a domain-centric and somewhat isolated fashion in sub-areas such as speech synthesis and recognition, language understanding, dialogue state tracking, turn taking and dialogue management, etc. This has resulted in limited dissemination of knowledge between these fields and while this phenomenon is being mitigated by the universality of deep learning underlying most recent advances in all of these areas, conversational AI platforms that provide proper infrastructure to support model training and evaluation can be a strong catalyst for research and development in the field. 

With advancements in joint and end-to-end learning \cite{joint_dilek, joint_rastogi, joint_zhao, e2e_bingliu, e2e_weston, e2e_moss, e2e_mila} along with widely used out of the box ASR and TTS solutions \cite{gcloud, polly, ms_speech}, several platforms and toolkits \cite[among others]{pydial,ParlAI,rasa,NeMo,convlab} have recently been proposed  for building Conversational AI systems. Each toolkit is largely designed around specific use-cases, with some being centered around research and others designed for scalability and use in production as described later. For this reason and others, for example unacceptable performance (e.g. occasional toxic or irrelevant output \cite{metoo_alexa, offensive_convai}) or lack of toolkits that can support both research and prototyping, there has been a disconnect between state-of-the-art research and its applications in the production systems. However, as conversational AI agents become more and more capable, new toolkits are needed that can bridge the gap between research and production and support quick prototyping of full conversational experiences with ASR, LU, DST/DM, LG, and TTS capabilities. There are only a few toolkits such as RASA \cite{rasa}, which  have been recently proposed and support both modeling and prototyping of an end-to-end experience. Most existing toolkits are centered around specific modules (e.g. LU) at the expense of other components (e.g. LG). Section \ref{sec:existing_toolkits} provides an overview of existing state-of-the-art toolkits. 

To address the challenges mentioned above, we propose the Plato Research Dialogue System \footnote{\url{https://github.com/uber-research/plato-research-dialogue-system}}, a flexible conversational AI research, experimentation, and prototyping platform. By following theoretical and well founded aspects of Spoken Dialogue Systems, Plato is designed in a manner which makes it easy to understand for people with different levels of expertise in conversational AI, researchers and developers. Plato's architecture and component-driven design supports statistical dialogue systems, multi-agent training (e.g. learning via agent-to-agent interaction\cite{georgila2014single, papangelis-etal-2019-collaborative, liu2017iterative, liu2018adversarial}), joint learning of modules \cite{joint_dilek, joint_rastogi, joint_zhao}, and end-to-end learning of conversational agents \cite{e2e_bingliu, e2e_weston, e2e_moss, e2e_mila}, while being flexible enough for quick prototyping and demonstration system development. Plato is agnostic to statistical learning frameworks, and developers and researchers can use any Python library (e.g. any Deep Learning or Reinforcement Learning library, etc.) they choose. As such, Plato seamlessly works with Ludwig \cite{ludwig} for code-less training of Deep Learning models and quick prototyping. 

Plato is designed to be modular, i.e. each conversational agent is composed of a number of modules that can run sequentially, in parallel, or in any combination of sequential and parallel. Each module can be a ``standard" component such as LU, DM, LG, or anything that fits the application's purpose (a part of speech tagger, a topic classifier, a joint model, etc.). This modular design helps make Plato more extensible and scalable, and allows several developers to build and experiment with individual modules simultaneously. Internally, each component of a conversational agent can be anything from a statistical model (trained online or offline) to a set of rules (e.g. using pattern matching for LU or templates for LG). Moreover, each component can call an API or service such as Google Cloud \cite{gcloud}, Amazon Transcribe \cite{aws_transcribe}, or Polly \cite{polly} for speech recognition, speech synthesis, or any other function. Besides building full conversational AI applications, Plato can be used to evaluate and experiment with various kinds of Natural Language Processing (NLP) tasks such as Sentiment Analysis, Topic Modeling, Dialogue State Tracking, Social Language Generation, and others.

\subsection{Existing Publicly Available Conversational AI Systems} \label{sec:existing_toolkits}

Several toolkits and conversational AI platforms have been proposed recently. The following are some state-of-the-art and widely adopted platforms: 

\begin{itemize}

\item \textbf{PyDial} \cite{pydial} is a toolkit for statistical modeling of multi-domain Dialogue Systems that supports Reinforcement Learning (RL) and Deep RL models. It is modular and supports customization of modules, however it is primarily designed for research rather than production and therefore requires deep knowledge of the conversational AI field.

\item \textbf{ParlAI} \cite{ParlAI} is a dialogue research framework which contains popular baselines and state-of-the-art models corresponding to a variety of tasks such as SQuAD \cite{squad}, bAbi \cite{bAbi}, and visual question answering. It provides integration with Amazon Mechanical Turk \cite{amt} for data collection and with Facebook Messenger for applications. Although ParlAI is rich in terms of models and tasks, it is not modular and the design is not centered around standard Dialogue System architectures, containing modules such as Language Understanding, Policy, Dialogue State Tracking and Language Generation, and therefore custom code may be required to build each component for a new task. Similar to ParlAI, \textbf{DeepPavlov} \cite{DeepPavlov} provides models (and corresponding code) for various tasks. However, the agent class provided in both frameworks needs extensions in order to be applicable to new tasks.

%\item OpenDial \cite{opendial} is a Java based modular library which supports Language Understanding, Dialogue Management, and Language Generation. Since it is written in Java, it is limited in terms of Deep Learning capabilities and using other related libraries. Furthermore, it does not have state of the art models implemented and support rule based modules.

\item \textbf{Rasa} \cite{rasa} is a rich set of libraries for building Dialogue Systems. It is developer-friendly and supports training of LU and dialogue policy (action) models with very little code. Rasa also supports model-based dialogue policies and provides wrappers to easily train various models through configuration files (similar to Plato). Rasa supports online learning, that is users can interact with the agent to assess its behaviour and for other debugging purposes. It also supports learning conversations and corresponding actions through examples by defining the conversations in the form of stories. Although Rasa is quite robust in terms of functionality, its architecture is quite different from standard module-based Dialogue System architectures. That being said, custom code can be written along with existing modules (such as the Agent) to extend functionalities of Rasa. On the other hand, even though Rasa provides functionalities to experiment with various state-of-the-art models, it is primarily designed for scalability.  

%\item ReAgent

\item \textbf{Neural Modules (NeMo)} \cite{NeMo} is a toolkit supporting speech recognition and NLP models such as named entity recognition, and intent and slot filling. Even though NeMo has integrations with state-of-the-art models such as BERT \cite{bert} and Transformers \cite{transformers}, it currently does not support all the building blocks of a Dialogue System. Several modules such as LG, policy, DST, and DM are missing. Hence, it does not provide an end-to-end conversational experience. 

\item \textbf{ConvLab} \cite{convlab} is a multi-domain dialogue research platform whose objective is to enable researchers to quickly set up experiments with reusable components and modules to compare across different approaches. ConvLab provides evaluation modules (leveraging humans or algorithms) for comparing and evaluating different policies or models. Even though it supports most standard dialogue system components (except ASR and TTS modules), ConvLab is primarily designed for research and evaluation of various state-of-the-art implementations.

\end{itemize}

Other frameworks such as SimDial \cite{simdial}, OpenDial \cite{opendial}, and Olympus \cite{olympus} (among others) seem to be less active in terms of support for state of the art models and learning frameworks when compared to the aforementioned toolkits.

The goal we try to achieve with Plato, is to develop a conversational AI platform that is one abstraction level above the aforementioned platforms\footnote{Therefore, each of the use cases supported by these platforms can be supported by Plato.}, is easy to understand and debug and that is agnostic to the underlying libraries that implement each component's statistical models or APIs. Plato can support any kind of Python module such as PyTorch \cite{pytorch}, TensorFlow \cite{tf}, or Keras \cite{keras}, and can integrate with code-free Deep Learning libraries such as Ludwig \cite{ludwig}, so that users at any level of experience can build a quick prototype and experiment with state-of-the-art ideas, while expert users can dive deeper and build custom and more scalable systems. Plato supports continuous and code-free training of each component and can be used to implement agent architectures that consist of a single module (e.g. an end-to-end network) or many serial or parallel modules. Several Deep Learning and Reinforcement Learning based examples are provided, as well as integrations with standard datasets such as DSTC \cite{dstc} and MetalWoz \cite{metalwoz} to guide the users in leveraging the full capabilities of Plato. Lastly, Plato supports multi-agent interaction, where several agents can interact with each other and train any of their components online.

\begin{figure}
    \centering
    \includegraphics[scale=0.3]{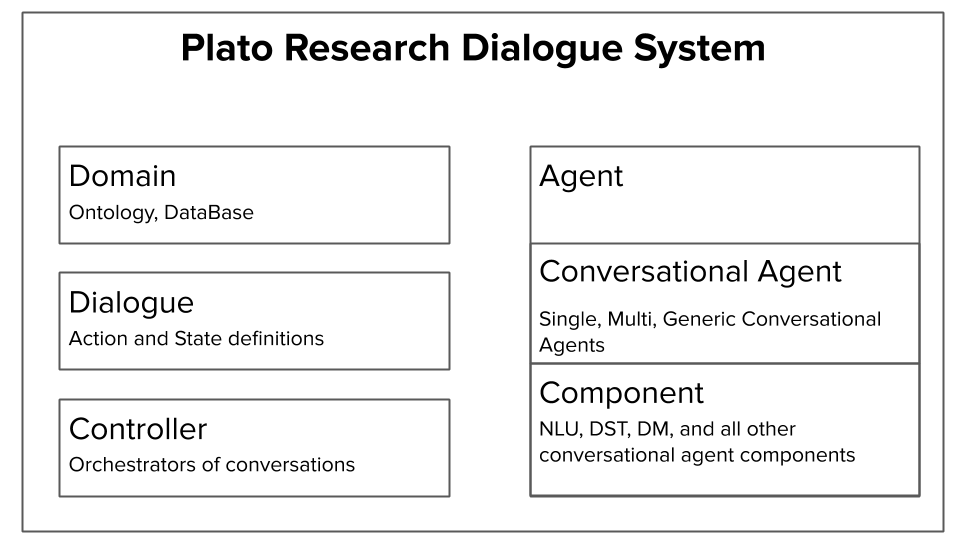}
    \caption{Major components of Plato: Domain, Dialogue, Controller, and Agent.}
    \label{fig:overview}
\end{figure}

\begin{figure}
  \centering
  \begin{tabular}[b]{c}
    \includegraphics[width=.51\linewidth]{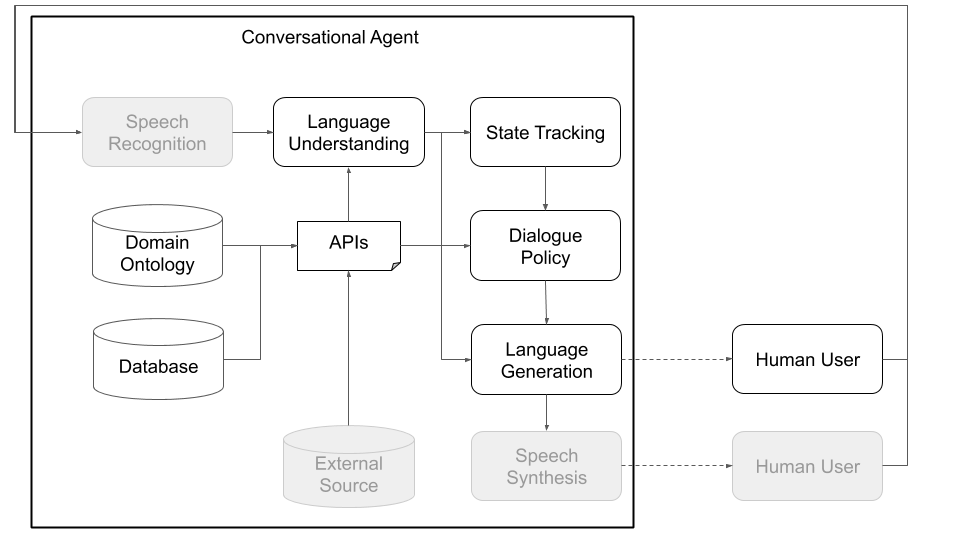} \\
    \small (a)
  \end{tabular} \qquad
  \begin{tabular}[b]{c}
    \includegraphics[width=.39\linewidth]{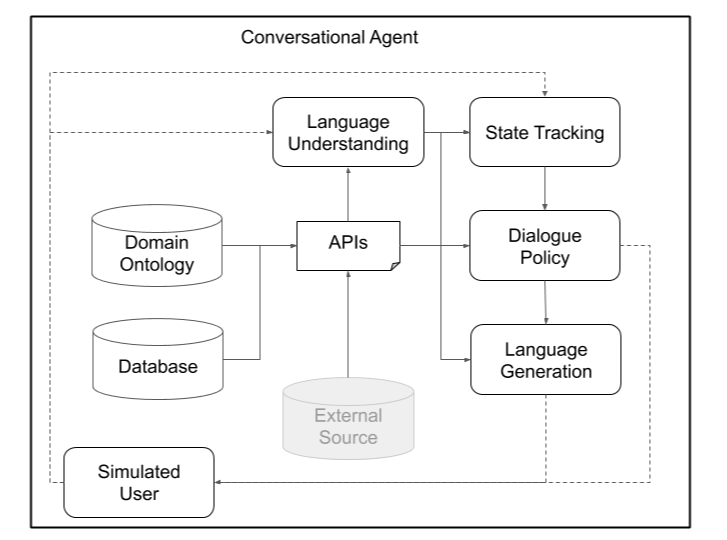} \\
    \small (b)
  \end{tabular}
  \caption{(a) Plato with humans. (b) Plato with Simulated Users.}
  \label{fig:plato_arch}
\end{figure}

\section{Plato Architecture}
Plato Research Dialogue System is a platform that can be used to create, train, and evaluate conversational AI agents. It has four main components, namely: 1) \textbf{dialogue} which defines and implements dialogue acts and dialogue states; 2) \textbf{domain} which includes the ontology of the dialogue and the database that the dialogue system queries; 3) \textbf{controller} which  orchestrates the conversations; and 4) \textbf{agent} which implements different components of each conversational agent.  These four major components are shown in Figure \ref{fig:overview} and are described in detail in the following sections. In Plato, each of these components is instantiated using configurations that are described in a YAML file (see Appendix 5 for an example).

\begin{figure}
    \centering
    \includegraphics[scale=0.3]{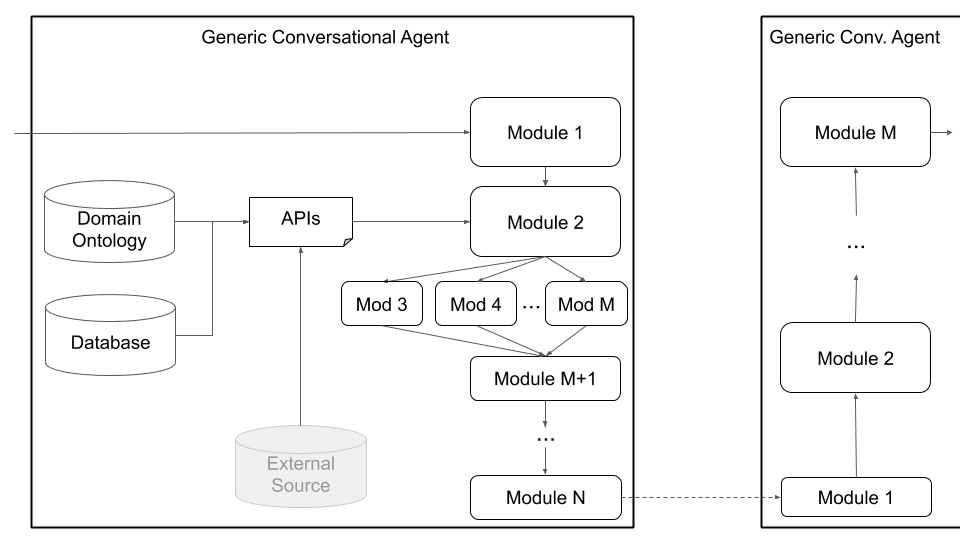}
    \caption{Communication between two  Plato Generic Conversational Agents. Each module performs a function such as joint language understanding and state tracking.}
    \label{fig:generic_agent_arch}
\end{figure}

\begin{figure}
    \centering
    \includegraphics[scale=0.3]{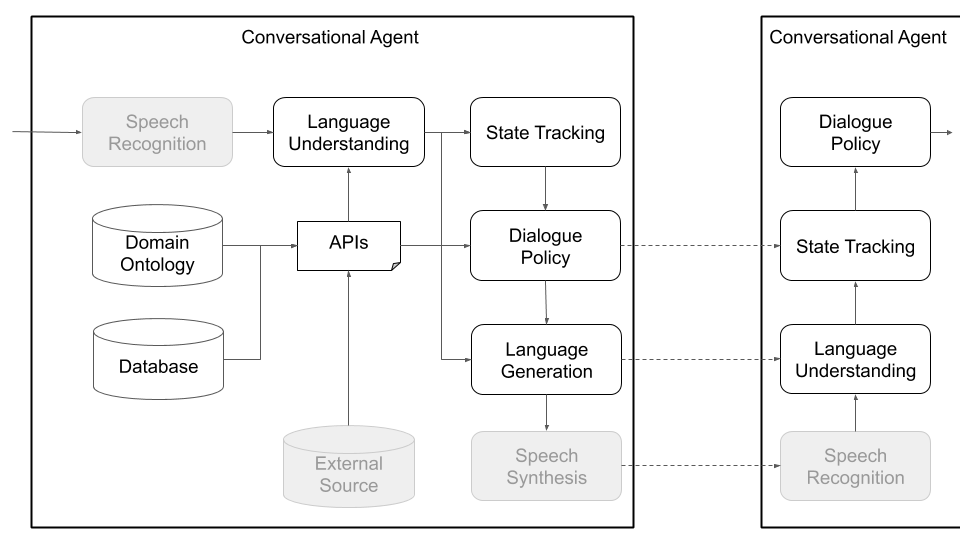}
    \caption{Communication between two Plato Conversational Agents. Grayed out components are external to Plato. The Agents can communicate via structured information (sets of Dialogue Acts: Dialogue Policy to State Tracking), via text (Language Generation to Language Understanding), or via Speech (Speech Synthesis to Speech Recognition).}
    \label{fig:multi_agent_arch}
\end{figure}

Plato is designed to be modular and flexible and supports standard conversational agents, such as the ones depicted in Figure \ref{fig:plato_arch}, as well any customized conversational agent architecture, for example joint LU and DST via the \emph{generic conversational agent} that operates as shown in Figure \ref{fig:generic_agent_arch}. At the highest level of abstraction, Plato supports agents communicating with other agents while adhering to some dialogue principles (Figure \ref{fig:multi_agent_arch}). These principles define \emph{what each agent can understand} (an ontology of entities, or meanings e.g. price, location, user preferences, cuisine types) and \emph{what it can do} (ask for more information, provide some information, call an API, etc.). The agents can communicate over speech, text, or structured information (e.g. dialogue acts) and each agent has its own configuration. In the rest of this section we will go over the major components of Plato.

\subsection{Dialogue}
Plato facilitates conversations between agents via well-defined concepts in dialogue theory, such as dialogue states and dialogue acts. A Plato agent, however, may need to perform actions not directly related to dialogue (e.g. call an API) or actions that communicate information in modalities other than speech (e.g. show an image). Therefore, Plato models Actions and States as abstract containers out of which Dialogue Acts and Dialogue States are created. If needed for specific applications (e.g. multi-modal conversational agents) we may have task-specific Dialogue Acts and States. 

\subsection{Domain}
For implementing a slot-filling \footnote{Slot-filling systems are dialogue systems whose primary purpose is to provide a natural language interface to an API, for example flight booking, information retrieval, etc.} task-oriented dialogue system in Plato, we need to specify two elements that constitute the domain of the dialogue system:

\begin{enumerate}
    \item {\bf Ontology} of the dialogue system. In task-oriented applications, the ontology determines informable slots (user provides that information), requestable slots (user requests that information), and system requestable slots (system requests that information) for the conversation, thereby reflecting the schema of the database that the agent queries to get the right information to be sent out. 
    
    \item {\bf Database} of items (restaurants, dishes, answers to questions, etc.). While the database could already exist, Plato provides utilities to construct the domain and the database of a dialogue system from data.
\end{enumerate}

For instance, in the case of a conversational agent for restaurant reservation, the cuisine could be thought of as an informable slot and the database could contain information about restaurants of different cuisines, their price range, address, etc. For non-slot-filling applications, Plato ontologies and databases can be extended to meet task- or domain-specific requirements.

\paragraph{Domain creation} Plato provides a utility that makes it easy to generate an ontology (a \code{.json} file) and a database (SQLite) from a \code{.csv} file, with columns representing item attributes and rows representing items (for an example, see \code{plato/example/data/flowershop.csv}). The main purpose of this utility is to help quick prototyping of conversational agents. The command \code{plato domain --config <PATH/TO/CONFIG.YAML>} calls the utility and generates the appropriate \code{.db} and \code{.json} files that define the domain. In the YAML configuration file, the user specifies details such as the path to the input \code{.csv} file, the columns that represent informable slots, etc. 

\subsection{Controller}
In Plato, controllers are objects that orchestrate the conversations between the agents. A controller instantiates the agents, initializes them for each dialogue, passes input and output appropriately, and keeps track of statistics. Note that it is each agent's responsibility to interact with the world if needed (listen to the microphone, call an API, etc.).

\begin{figure}[h]
    \centering
    \includegraphics[scale=0.4]{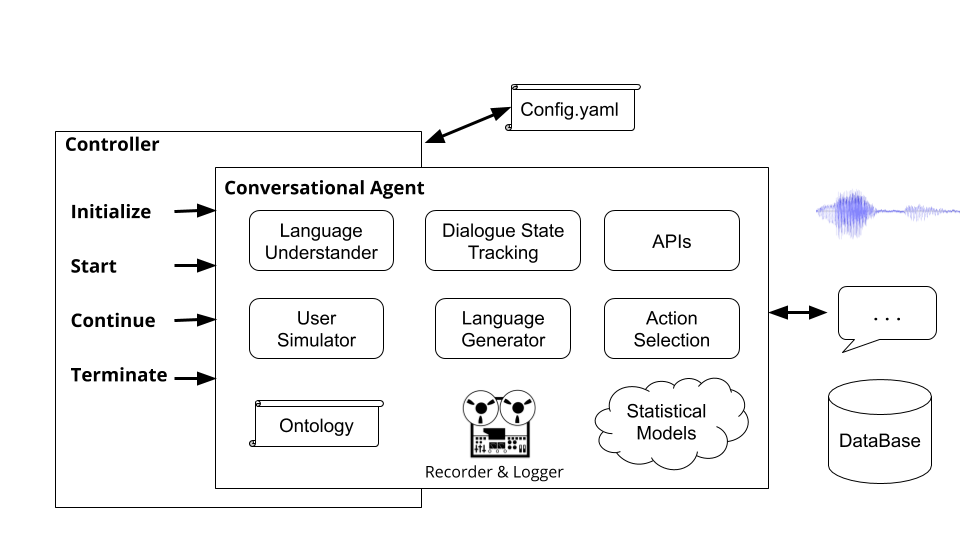}
    \caption{Main functions of Plato's basic controller.}
    \label{fig:controller}
\end{figure}

Running the command \code{plato run -{}-config <PATH/TO/CONFIG.YAML>} runs Plato's basic controller, shown in Figure \ref{fig:controller}. This command receives a value for the \code{-{}-config} argument which points to a Plato application configuration YAML file. In this configuration file, details of the agent(s) involved in the dialogue, paths to the ontology and database of the dialogue, along with other parameters and options are specified.

Plato also has a controller that comes with a Graphical User Interface (GUI). This controller can be started by running \code{plato gui -{}-config <PATH/TO/CONFIG.YAML>}. This controller is very similar to Plato's basic controller except that the user is prompted through a GUI as opposed to interacting via the terminal.

\subsection{Agent}
\begin{figure}[h]
    \centering
    \includegraphics[scale=0.5]{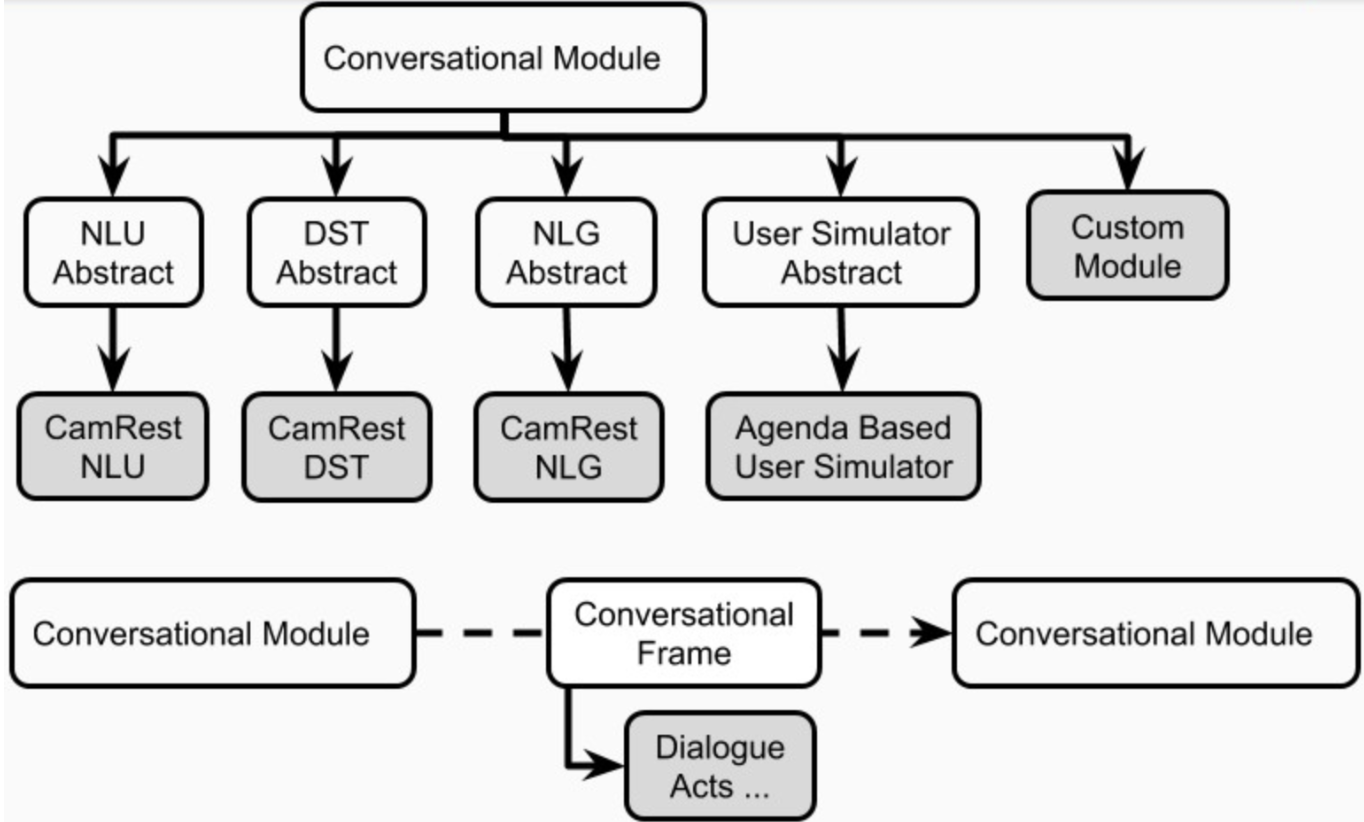}
    \caption{Agent components in Plato inherit from Conversational Module, which facilitates communication with other components via Conversational Frames.}
    \label{fig:plato_components}
\end{figure}

Every conversational application in Plato could have one or more agents. Each agent has a role (assistant, user, tourist, tutor, etc) and a set of components such as  NLU, DM, DST, dialogue policy, and NLG \footnote{Plato supports external API calls for automatic speech recognition (ASR) and text to speech synthesis (TTS).}. An agent could have one explicit module for each one of these components or alternatively, some of these components could be combined into one or more modules (e.g. joint or end-to-end agents) that can run sequentially or in parallel (Figure \ref{fig:generic_agent_arch}). All components inherit from \code{conversational\_module}, as shown in Figure \ref{fig:plato_components}, which facilitates communication between any component via \code{conversational\_frames}.
Each one of these modules could be either rule-based or trained. In the next subsections, we will describe how to build rule-based and trained modules for agents.

\subsubsection{Rule-based modules}
Plato provides rule-based versions of all components of a slot-filling conversational agent: \code{slot\_filling\_nlu}, \code{slot\_filling\_dst}, \code{slot\_filling\_policy}, \code{slot\_filling\_nlg}, and the default version of the Agenda-Based User Simulator \cite{schatzmann2007agenda} \code{agenda\_based\_us}. These can be used for quick prototyping, baselines, or sanity checks. Specifically, all of these components follow rules or patterns conditioned on the given ontology and sometimes on the given database and should be treated as the most basic version of what each component should do.

\subsubsection{Trained modules}
Plato supports training of agents’ components in an online (during the interaction) or offline (from data) manner, using any machine learning framework. Virtually any model can be loaded into Plato as long as Plato’s Input/Output interface is respected. For example, if a model is a custom NLU, it simply needs to inherit from Plato's NLU abstract class (\code{plato.agent.component.nlu}), implement the necessary functions and pack/unpack the data into and out of the custom model.

\paragraph{Plato internal experience}
To facilitate online learning, debugging, and evaluation, Plato keeps track of its internal experience using a utility called the \code{dialogue\_episode\_recorder}, which stores information about previous dialogue states, actions taken, current dialogue states, utterances received and utterances produced, rewards received, and a few other constructs including a custom field that can be used to track anything else that cannot be contained by the aforementioned categories. At the end of a dialogue or at specified intervals, each conversational agent will call the \code{train()} function of each of its internal components, passing the dialogue experience as training data. Each component then picks the parts it needs for training.

To use learning algorithms that are implemented inside Plato, any external data, such as DSTC2 data, should be parsed into this Plato experience so that they may be loaded and used by the components currently under training. Alternatively, users may parse the data and train their models outside of Plato and then load the trained model when they want to use it for a Plato agent.

\paragraph{Parsing data with Plato}
Plato provides parsers for DSTC2, MetaLWOZ, and Taskmaster data sets. These parsers can be used to create training data for different components of the agent based on these well known data sets. For other data sets the user should implement custom parsers to convert the data into Plato-readable format. Users can then load the data into Plato (via options in the configuration file) and train or fine-tune the desired components.

\paragraph{Training components of conversational agents}
There are two main ways to train each component of a Plato agent: online, as the agent interacts with other agents, simulators, or users and offline, from data. For online training users can determine the train schedule via hyper-parameters (train after how many dialogues, for how many epochs, how large is the experience pool and the minibatch, etc.). Moreover, users can use algorithms implemented in Plato or external frameworks such as TensorFlow, PyTorch, Keras, Ludwig, etc.  

\paragraph{Model Training with Plato} Besides supervised models, Plato also provides some implementations of reinforcement learning algorithms, such as Q-Learning or REINFORCE. Such algorithms can be used to train for example the agent's dialogue policy online, as the agent interacts with its environment. Plato provides the flexibility to train after each dialogue or at certain intervals. Note that while the reinforcement learning implementations are geared towards dialogue policies, Plato and Ludwig support online training of any component.

\paragraph{Training with Ludwig} Although virtually any modeling framework could be used in Plato, for building and training deep learning models for different components of conversational agents, Ludwig is a good choice when the aim is quick prototyping or for educational purposes, as Ludwig allows users to train models without writing any code. Users need to parse their data into .csv files, create a Ludwig configuration YAML file that describes the architecture, which features to use from the .csv and other parameters and then run a command in a terminal. This allows Plato to integrate with Ludwig models, i.e. load or save the models, train and query them. The trained models could be loaded into the modules through configuration files. In the tutorial of Plato we provide examples of building and training language understanding, generation, dialogue policy, and dialogue state tracking models for Plato using Ludwig.

\paragraph{Training with other frameworks} To use other learning frameworks (TensorFlow, PyTorch, Keras, etc.) users simply need to write a class that interfaces with the trained model (i.e. parses the input Plato provides and processes the model's response into structures that Plato understands).

\begin{comment}
\section{Plato Architecture}
\begin{itemize}
    \item \color{RubineRed}Explain configuration files managing different parts of Plato

\end{itemize}
    \subsection{Dialog}
    \subsection{Controller}
        \begin{itemize}
            \item ...
            \item GUI
        \end{itemize}
\subsection{Domain}
\subsection{Agent}
    
In this section, we describe Plato's core components.

        \begin{itemize}
            \item What are agent components
            \item ...
        \end{itemize}
        \subsubsection{Standard Components}
        \subsubsection{{Custom Components}}
        
\end{comment}

\section{Plato Settings}
As mentioned in the previous section, Plato supports interaction between a conversational agent and human agents, simulated users, or other conversational agents. In this section we discuss the details of each configuration. 

\subsection{Single Agent}
A single Plato conversational agent can interact with a) human users, via text or speech, command line or graphical interface; b) simulated users, via dialogue acts or text; and c) data (loading data into Plato in order to train components or generating simulated data). The specifics of each interaction are specified in YAML configuration files, which have three main sections:

\begin{itemize}
    \item {\bf GENERAL} defines the mode of the interaction (e.g. with a simulator, via speech or text, etc), the number of agents to spawn, paths to experience logs, and global arguments (passed to all components of each agent, for convenience).
    \item {\bf DIALOGUE} defines dialogue-specific settings, such as number of dialogues to run for, domain, etc.
    \item {\bf AGENT\_i} defines each agent's settings, such as role, and component-specific settings, such as model paths, learning rates, or other arguments each component needs. Note that the global arguments from the GENERAL section are also passed to each component of each agent.
\end{itemize}

For an example, see Appendix A.5 or in the Plato codebase:\\ \code{plato/example/config/application/CamRest\_user\_simulator.yaml}

\subsection{Multiple Agents}
The basic controller \code{plato.controller.basic\_controller} provides support for two agents interacting with each other. Similar to the single agent configuration, this can be done in dialogue acts or in free text. Interaction through speech is also possible of course, for example the agents can exchange \code{.wav} files instead of text. 

In this setting, it is possible to train all components of each agent on-line and concurrently or following any desired schedule, e.g. alternating training, training in batches, etc. These options can be defined in the yaml configuration file, which is very similar to the single agent case but defines multiple \code{AGENT} sections (see \code{plato/example/config/application/MultiAgent\_train.yaml}). Plato provides some example multi-agent reinforcement learning algorithm implementations, for concurrent dialogue policy learning.

Example use cases of multiple conversational agents include
\begin{itemize}
    \item {\bf General Sum Games} such as negotiations, where the agents' objectives are not completely aligned (but are not completely opposite either).
    \item {\bf Multi-party interactions}, where one or more conversational agent interact with groups of other agents, for example family dinner ordering, playing board games, etc.
    \item {\bf Smart Home}, where a conversational agent is a point of contact between human agents, conversational agents, and non-conversational agents.
\end{itemize}

To support complex use cases, a new controller may be necessary, to make sure that information is passed correctly between the agents.

\subsection{Graphical User Interface}
\begin{figure}[h]
    \centering
    \includegraphics[width=\linewidth]{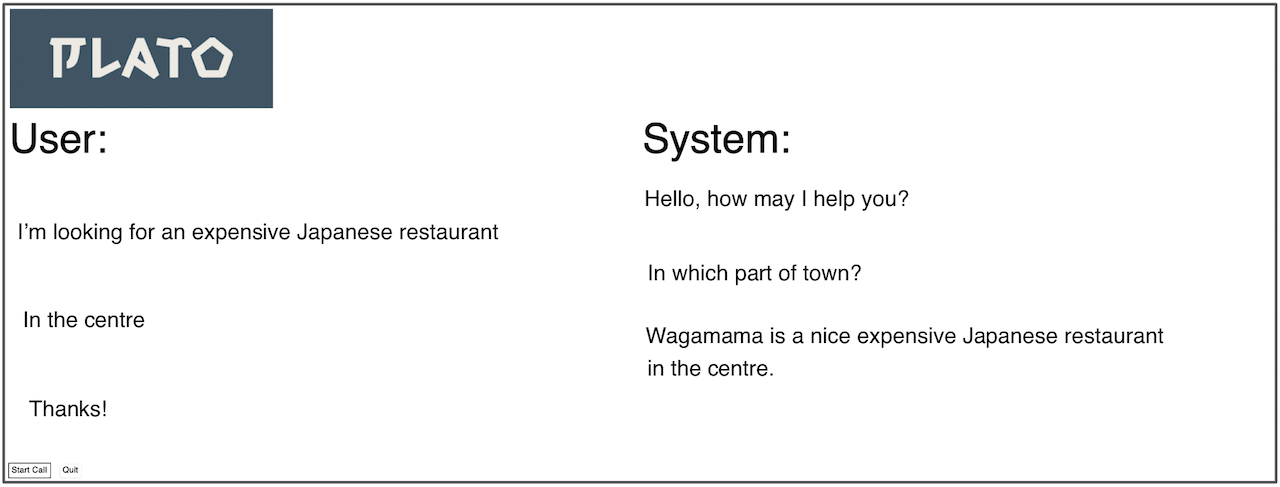}
    \caption{Plato's PySimpleGUI-based Graphical User Interface.}
    \label{fig:gui}
\end{figure}

Plato provides a dedicated controller to handle the GUI. In its current implementation it supports interaction between two agents, as shown in Figure \ref{fig:gui}. This interface is just an example based on PySimpleGUI, a flexible and easy to use package.

\section{Conclusion}
We have introduced the Plato Research Dialogue System, a flexible platform for research and development of conversational AI agents. With an easy-to-understand extensible design, Plato provides the  infrastructure required by virtually any conversational AI agent architecture and supports any Python machine learning framework for the agent's components. As Plato continues to grow, more models, algorithms, and metrics will be integrated as well as more examples, tutorials and data parsers to publicly available datasets. 

Plato can be obtained from: \url{https://github.com/uber-research/plato-research-dialogue-system}

\section{Acknowledgements}
We would like to thank Michael Pearce and Zack Kaden for their contributions.

\bibliographystyle{unsrt} 
\bibliography{references}

\appendix
\section{Appendix}
Here we provide some examples of how to run Plato. This appendix assumes some general knowledge of Spoken Dialogue Systems theory, prominent data sets, benchmarks, etc. A full user guide and installation instructions can be found at \url{https://github.com/uber-research/plato-research-dialogue-system}. 

\subsection{Quick Start}
To run Plato after installation, you can simply run the \code{plato} command which receives 4 sub-commands:
\begin{itemize}
    \item \code{plato run --config <CONFIG.YAML>}
    \item \code{plato gui --config <CONFIG.YAML>}
    \item \code{plato domain --config <CONFIG.YAML>}
    \item \code{plato parse --config <CONFIG.YAML>}
\end{itemize}

Each of these sub-commands receives a value for the \code{--config} argument that points to a configuration file. For some quick examples, try the following configuration files for the Cambridge Restaurants domain:

\begin{itemize}
    \item \code{plato run --config CamRest\_user\_simulator.yaml}
    \item \code{plato run --config CamRest\_text.yaml}
    \item \code{plato run --config CamRest\_speech.yaml}
    \item \code{plato gui --config CamRest\_GUI\_speech.yaml}
\end{itemize}

\subsection{Train a module}
There are several options for training a module in Plato as detailed in the user guide\footnote{\url{https://github.com/uber-research/plato-research-dialogue-system/README.md}}. Plato provides support for offline and online training (allowing for custom training schedules). For online training, users need to implement a \code{train()} function in their module (called according to the schedule) which may directly train the custom model, call an API for training it, etc. Offline training can happen inside or outside of Plato. In this section, we will demonstrate the latter, as training within Plato is relatively straightforward.

In this example, we show how to train an NLU using DSTC2 \cite{dstc} data. Our model will jointly predict the intent and Begin-In-Out tags. Table \ref{tab:nlu_train}, below, shows a snapshot of the training data.

\begin{table}[h]
\begin{tabular}{l|c|l}
     {\bf transcript} & {\bf intents} & {\bf BIO tags} \\
     \hline
    expensive restaurant that serves vegetarian food & inform & B-inform-pricerange O O O B-inform-food O \\
    asian oriental type of food & inform & B-inform-food I-inform-food O O O \\
    what is the phone number & request\_phone & O O O O O \\
    thank you good bye & bye thankyou & O O O O \\
    how about french food & reqalts inform & O O B-inform-food O
\end{tabular}
\vspace{2mm}
\caption{Sample NLU training data for DSTC2.}
\label{tab:nlu_train}
\end{table}

Using Ludwig, we can define our model with a simple configuration file (which can be found in Plato's user guide) and call the following command to train the model (some details are omitted, see the full Plato guide):

\code{ludwig experiment --model\_definition\_file ludwig\_config.yaml --data\_csv NLU.csv}

To load the trained model in Plato a wrapper class needs to be written to interface with the specific model, i.e. to properly format the input, query the model, and properly parse its output so that downstream Plato components can understand it. Specifically for Ludwig, these classes are provided in Plato. The final step then is to modify Plato's configuration file to point to the Ludwig-based NLU and to call plato using that configuration.

\subsection{Load pre-trained models}
Large pre-trained models from libraries such as Huggingface \cite{wolf2019transformers} are becoming a standard in Convsational AI. Plato inherently supports such models as from Plato's perspective they are no different than any other statistical model. For example, to use a pre-trained Huggingface BERT model for NLU, one needs to write a class implementing Plato's NLU interface and make sure to load and query BERT appropriately (see \url{https://github.com/huggingface/transformers} for detailed instructions and documentation for Huggingface models and see Plato's user guide for a relevant tutorial).

\subsection{Other topics}
Besides the above use cases, Plato provides a variety of utilities such as data parsing, database creation, and simulated data generation. For more details please see the full user guide.

\subsection{Example YAML configuration file}
\begin{figure}[h]
    \centering
    \includegraphics[width=0.75\linewidth]{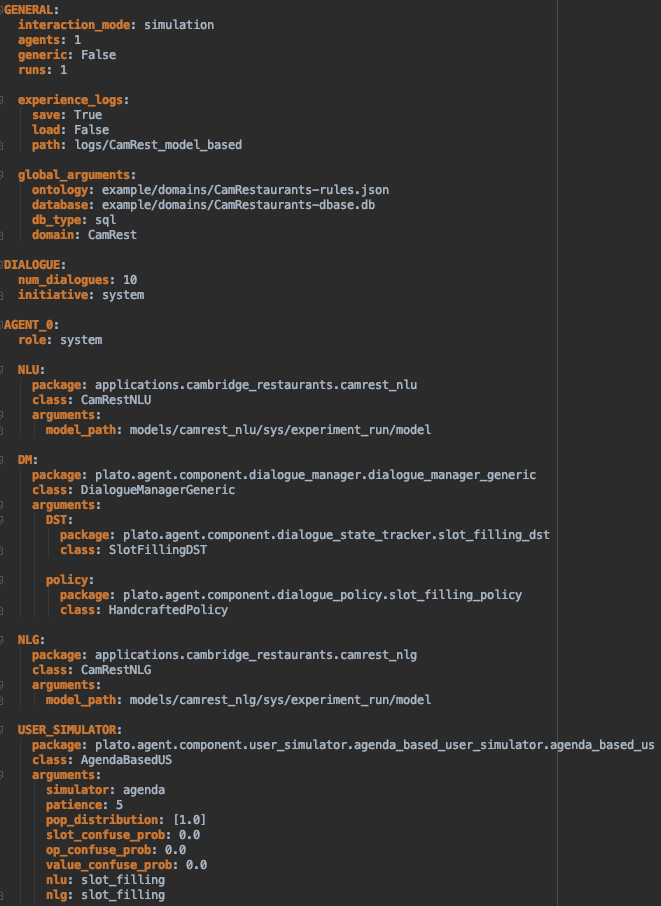}
    \caption{Example Plato YAML configuration file.}
    \label{fig:yaml}
\end{figure}

% \printbibliography
\end{document}